\pdfoutput=1
\documentclass[aps,prb,reprint,superscriptaddress,showpacs,amsmath,floatfix]{revtex4-1}

\usepackage{graphicx}

\begin{document}


\title{Magnetism dependent phonon anomaly in LaFeAsO observed via inelastic 
x-ray scattering }


\author{S. E. Hahn}
\affiliation{Department of Physics and Astronomy, Iowa State University, Ames, IA, 50011, USA}
\affiliation{Division of Materials Science and Engineering, Ames Laboratory US-DOE, Iowa State University, Ames, IA 50011, USA}
\author{G. S. Tucker}
\affiliation{Department of Physics and Astronomy, Iowa State University, Ames, IA, 50011, USA}
\affiliation{Division of Materials Science and Engineering, Ames Laboratory US-DOE, Iowa State University, Ames, IA 50011, USA}
\author{J.-Q. Yan}
\affiliation{Materials Science and Technology Division, Oak Ridge National Laboratory, Oak Ridge, TN, 37831, USA}
\affiliation{Department of Materials and Engineering, The University of Tennessee, Knoxville, TN, 37996, USA }
\author {A. H. Said}
\affiliation{ Advanced Photon Source, Argonne National Laboratory, Argonne, IL 60439, USA}
\author{B. M. Leu}
\affiliation{ Advanced Photon Source, Argonne National Laboratory, Argonne, IL 60439, USA}
\author {R. W. McCallum}
\affiliation{Division of Materials Science and Engineering, Ames Laboratory US-DOE, Iowa State University, Ames, IA 50011, USA}
\author{E. E. Alp}
\affiliation{ Advanced Photon Source, Argonne National Laboratory, Argonne, IL 60439, USA}
\author{T. A. Lograsso}
\affiliation{Division of Materials Science and Engineering, Ames Laboratory US-DOE, Iowa State University, Ames, IA 50011, USA}
\affiliation{Department of Materials Science and Engineering, Iowa State University, Ames, IA, 50011 USA}
\author{R. J. McQueeney}
\affiliation{Department of Physics and Astronomy, Iowa State University, Ames, IA, 50011, USA}
\affiliation{Division of Materials Science and Engineering, Ames Laboratory US-DOE, Iowa State University, Ames, IA 50011, USA}
\email[]{mcqueeney@ameslab.gov}
\author{B. N. Harmon}
\affiliation{Department of Physics and Astronomy, Iowa State University, Ames, IA, 50011, USA}
\affiliation{Division of Materials Science and Engineering, Ames Laboratory US-DOE, Iowa State University, Ames, IA 50011, USA}


\date{\today}

\begin{abstract}
The phonon dispersion was measured at room temperature along (0,0,L) in the tetragonal phase of LaFeAsO using inelastic x-ray scattering.  Spin-polarized first-principles calculations imposing various types of antiferromagnetic order are in better agreement with the experimental results than nonmagnetic calculations, although the measurements were made well above the magnetic ordering temperature, $T_N$. Splitting observed between two $A_{1g}$ phonon modes at 22 and 26 meV is only observed in spin-polarized calculations. Magneto-structural effects similar to those observed in the AFe$_2$As$_2$ materials are confirmed present in LaFeAsO.  The presence of Fe-spin is necessary to find reasonable agreement of the calculations with the measured spectrum well above $T_N$. On-site Fe and As force constants show significant softening compared to nonmagnetic calculations, however an investigation of the real-space force constants associates the magnetoelastic coupling with a complex renormalization instead of softening of a specific pairwise force.    
\end{abstract}


\maketitle

Despite rather convincing arguments that superconductivity in the AFe$_{2}$As$_{2}$ (A=Ca,Sr,Ba,Eu) and RFeAsO (R=La,Ce,Pr,Nd,Sm,Gd)-based compounds does not originate from conventional electron-phonon 
coupling,\cite{PRL.101.026403} these systems do display significant sensitivity to the lattice geometry. For example, the size of the Fe moment is sensitive to the lattice parameters and As position, as shown by Density functional theory (DFT) calculations.
One thus expects strong magneto-structural coupling in these compounds.\cite{PhysicaC.469.425} 
Also, measurements of the room temperature phonon density-of-states (DOS) in LaFeAsO indicated some 
disagreement with non-spin-polarized DFT calculations.\cite{PRL.101.157004,PRB.78.174507}  Distinct features in the 
phonon DOS, likely associated with atomic displacements in the Fe-As plane, were observed at 
significantly lower energies than non-magnetic calculations suggest. It was noted (empirically) 
that softening of the Fe-As force constants by 30\% brings the calculated phonon 
DOS into better agreement with the data.\cite{JPSJ.77.103715}  Theoretical studies have shown that strong coupling between Fe magnetism and the As  position leads to the softening of the Fe-As force constants, thereby explaining the observed phonon spectra.\cite{NatPhys.5.141}

While these magnetostructural effects are well documented in the AFe$_2$As$_2$-based 
systems, it is not clear if the same effects are present in the RFeAsO system. One key 
example of this coupling in CaFe$_{2}$As$_{2}$ comes from the observation of a transition from the 
antiferromagnetic state to a non-magnetic ``collapsed tetragonal" state 
 under applied pressure.\cite{PRB.78.184517}  In this case, 
a reduction of the \textit{c}-axis lattice parameter by 9.5\% is associated with the complete 
collapse of the Fe magnetic moment.\cite{PRB.79.060510}

The lattice vibrational frequencies associated with \textit{c}-axis vibrations
of Ca and As atoms in CaFe$_{2}$As$_{2}$ and BaFe$_{2}$As$_{2}$ 
\cite{PRB.79.220511,PRB.80.214534,JPCS.251.012008} 
have been shown by inelastic neutron and x-ray scattering to disagree with predictions of non-spin polarized DFT calculations. In particular, the energy splitting between c-axis phonon branches  containing As displacements was found to be in strong disagreement with non-spin polarized calculations.   
Ultimately, spin-polarized calculations in the local spin density approximation that include the 
AFM order present at lower temperatures were required to bring the calculated 
phonon dispersion into better agreement with room temperature measurements.\cite{PhysicaC.469.425}  Our group was able to confirm the role of 
magnetism in \textit{c}-axis polarized Ca and As modes in CaFe$_{2}$As$_{2}$ 
using single-crystal inelastic x-ray scattering (IXS) measurements at the Advanced Photon Source (APS) in combination 
with spin-polarized calculations using the Perdew-Burke-Ernzerhof (PBE) 
exchange-correlation functional.\cite{PRB.79.220511} 

It might be expected that the presence of RO layers, which results in a larger spacing of the 
FeAs layers along the \textit{c}-axis, might mitigate these effects to some degree.  However, the difficulty 
of synthesizing RFeAsO in single-crystalline form has prevented a quantitative confirmation of similar 
magneto-structural coupling across the AFe$_2$As$_2$ and RFeAsO systems. Recently single crystal samples of LaFeAsO have become
available. \cite{APL.95.222504} IXS phonon data was recently reported on PrFeAsO$_{0.9}$ \cite{JPSJ.77.103715} and SmFeAsO\cite{PRB.80.220504} 
single-crystals, however the dispersions were only measured along the (100) direction.
Also, the role of spin-phonon coupling could not be ascertained since results were compared 
only to non-spin-polarized LDA calculations. 

LaFeAsO single crystals were synthesized in an NaAs flux at ambient pressure as 
described elsewhere.\cite{APL.95.222504} Inelastic x-ray scattering measurements were 
performed on the HERIX instrument at sector 30-ID-C of the Advanced Photon Source at Argonne National Laboratory
with incident beam energy of 23.724 keV and with an energy resolution of 1.44 meV.\cite{JSR.18.492,JSR.18.605} 
Scattering is described in terms of the tetragonal P4/nmm unit cell 
where $\mathbf{Q} = \frac{2\pi}{a}\left( h\mathbf{i} + k \mathbf{j} \right) + \frac{2\pi}{c}l\mathbf{k}$. 
The vectors \textbf{i}, \textbf{j}, and \textbf{k} are the fundamental translation unit vectors 
in real space. Below $T_S$=156K, the sample transforms to an orthorhombic structure with space group Cmma.
\cite{NAT.453.899,PRL.101.077005} The relationship between the Miller indices in the tetragonal P4/nmm and orthorhombic 
Cmma phase are, $h=\left(H_o+K_o\right), k=\left(H_o-K_o\right)$, and $l=L_o$. Below the magnetic ordering temperature $T_N$=138K, the sample develops long-range spin-density wave (SDW) AFM order. 
The sample was mounted in the $\left(hhl\right)$ plane in a displex for low temperature studies, 
and the displex was attached to a 4-circle diffractometer. 

Based on previous studies of \textit{c}-axis 
polarized phonons in CaFe$_2$As$_2$, we focused our study on phonon branches along the 
$(0,0,8+\xi)$ direction in the Brillouin zone. In order to study the dispersion 
and potential line broadening of the phonon modes, the scans were fit to several peaks 
using a pseudo-Voigt line profile. The normalized pseudo-Voigt function is given in 
Eqn. \ref{eq: pseudo-Voigt}, where $f_{G}\left(x;\Gamma\right)$ and 
$f_{L}\left(x;\Gamma\right)$ are normalized Gaussian and Lorentzian functions 
respectively. The mixing parameter  $\eta= 0.74$, and resolution 
full-width-at-half-maximum (FWHM) $\Gamma=1.44$ meV was determined from 
fits to the elastic scattering width of Plexiglas. 

\begin{equation}
f_{pV}=\left(1-\eta\right)f_{G}\left(x;\Gamma\right)+\eta f_{L}\left(x;\Gamma\right)
\label{eq: pseudo-Voigt}
\end{equation}

\begin{figure}
\begin{tabular}{c}
\includegraphics[]{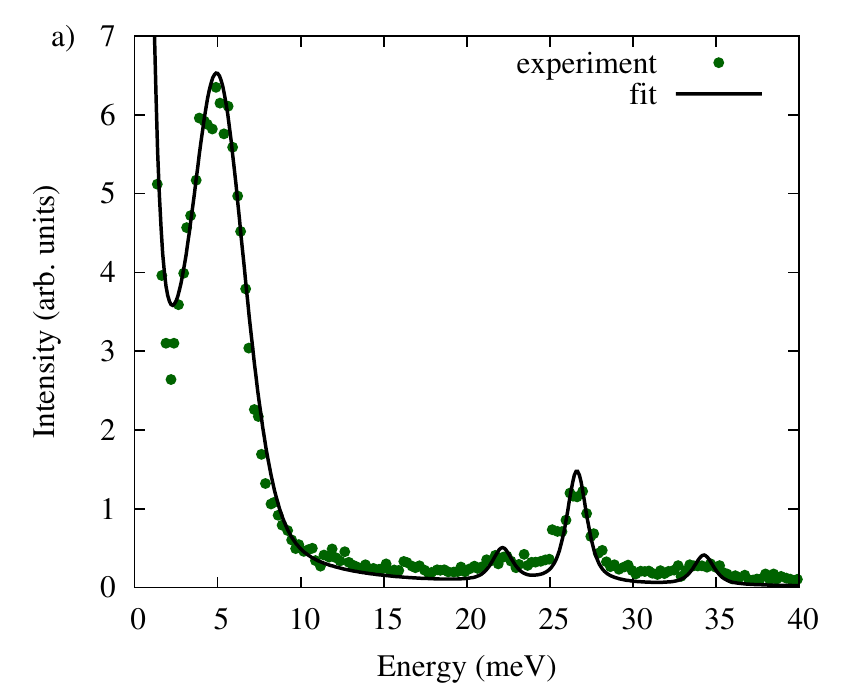}\tabularnewline
\includegraphics[]{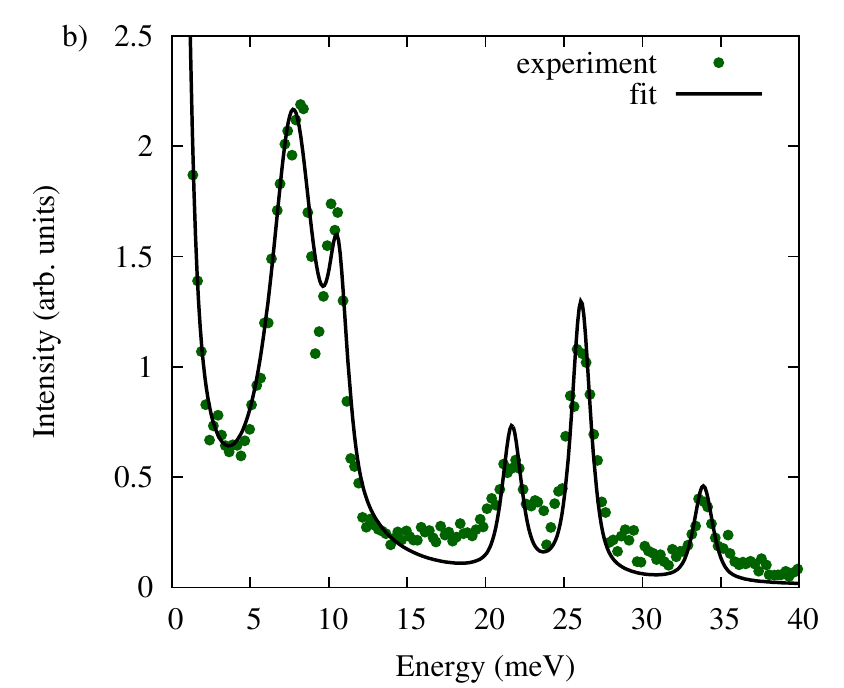}\tabularnewline
\end{tabular}
\caption{(color online) Energy scan at constant-Q at a) $Q=\left(0.0,0,8.3\right)$ and b) $Q=\left(0.0,0,8.5\right)$ measured at room temperature on LaFeAsO. Experimental data are given by solid green points. The black line is fit using a pseudo-Voigt function.\label{line scan}}
\end{figure}

\begin{figure}[htp]
\begin{tabular}{c}
\includegraphics[]{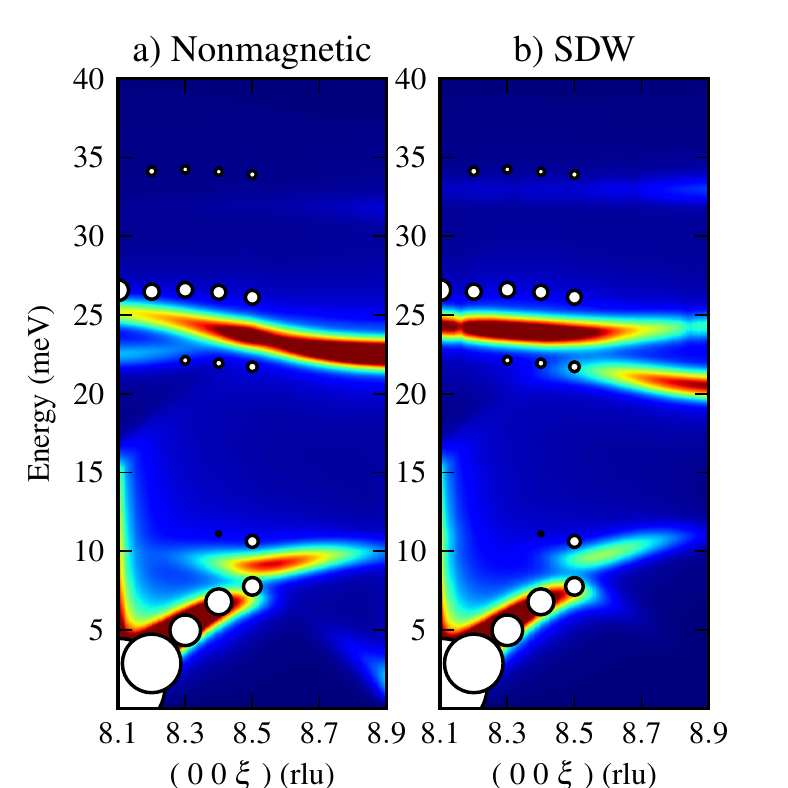}\tabularnewline
\includegraphics[]{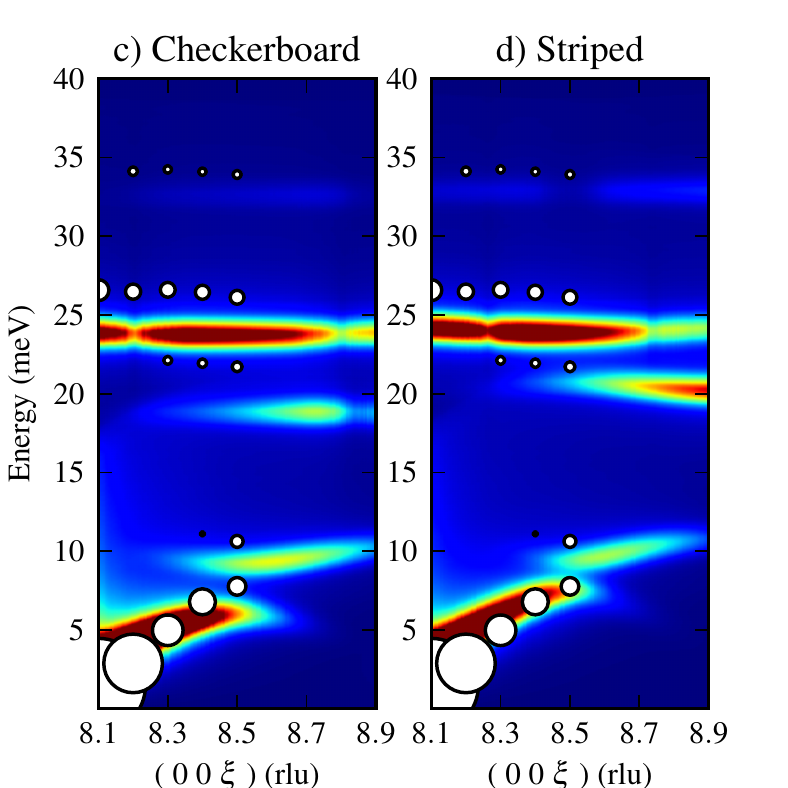}\tabularnewline
\end{tabular}
\caption{(color online) Contour plots of the calculated dynamical structure factor along (0,0,L). Values range from blue (no intensity) 
to red (high intensity), and have been multiplied by the energy to improve visibility of the optical modes. The white dots show the experimentally determined frequencies, as described in the text, with the intensity shown by the size of the dot. a) Nonmagnetic calculation b) SP calculation with SDW ordering, c) SP calculation with 
checkerboard ordering. d) SP calculation with striped ordering. \label{sf 00L}}
\end{figure}

Figure \ref{line scan} shows
a line scan consisting of several phonon excitations at $\mathbf{Q}=\left(0,0,8.3\right)$ and $\mathbf{Q}=\left(0,0,8.5\right)$ at room temperature. 
The peak positions for these and other scans were obtained from fits and used 
to construct the dispersion of phonon branches along the different scan directions, as shown in Fig. \ref{sf 00L}. 
The intensity of the phonon modes is also represented in Fig. \ref{sf 00L} 
by the diameter of the circles.

\begin{table}
\caption{\label{pos_mom}Theoretically relaxed and experimentally observed z-position for La and As atoms, and the associated 
magnetic moment per Fe atom and total energy. In each case the experimental lattice parameters of 
($a=4.03533 \text{\AA}$, $c=8.74090 \text{\AA}$) were used.\cite{jacs.130.3296,AiP.59.803}}
 \begin{ruledtabular}
 \begin{tabular}{l c c c c c}
& NM & SDW & Striped & Checkerboard & Exp.\cite{jacs.130.3296,AiP.59.803}\\
$z_{La}$& 0.13993 &  0.13875 & 0.13883 & 0.13887 & 0.14154 \\
$z_{As}$& 0.63829 &  0.64820 & 0.64770 & 0.64401 & 0.6512\\
$\mu_{Fe} $ & 0.0  &  2.32 & 2.30 & 1.91  & 0.36-0.78 \\
E(Ry) & 0.0 & -0.032 & -0.033 & -0.009 & \\
 \end{tabular}
 \end{ruledtabular}
\end{table}

In order to understand the features of the phonon dispersion, the
experimental measurements were compared to $\mathit{ab\ }initio$
calculations of the phonons. The phonon dispersion was calculated
using DFT and Density Functional Perturbation 
Theory (DFPT).\cite{QE-2009} There are significant differences in the experimental lattice 
parameters and parameters from the ``relaxed" structure with the lowest calculated energy. Also, in spin-polarized 
calculations with the experimentally observed AFM order, the lattice distorts into the 
orthorhombic Cmma structure observed experimentally at lower temperatures. With these difficulties in mind,  
the experimental lattice parameters at room temperature in the tetragonal phase ($a=4.03533 \text{\AA}$ , $c=8.74090 \text{\AA}$) 
were used for all calculations.\cite{jacs.130.3296,AiP.59.803} In addition, there is debate over the appropriate internal \textit{z}-parameter to 
use for the position of lanthanum  and arsenic atoms.\cite{PRL.102.217001,PRB.79.220511,PRB.81.144502,PhysicaC.469.425} For better accuracy of the calculated phonons, we chose the calculated 
relaxed positions where all forces were zero. Structural parameters used for the non-magnetic and 
spin-polarized calculations as well as experimental measurements are given in 
table \ref{pos_mom}. The pseudopotentials chosen used the Perdew-Burke-Ernzerhof (PBE) 
exchange correlation functional.\cite{pseudo,PhysRevLett.77.3865} Settings of an 8x8x4 (nonmagnetic),
4x4x4 (striped \& SDW) and 8x8x2 (checkerboard) k-mesh and 50 Ry and 660 Ry energy 
cutoffs for the wavefunctions and charge density were chosen to ensure meV precision of the calculated phonon dispersion. These parameters are similar to 
other phonon calculations of LaFeAsO.\cite{PhysRevLett.100.237003, PhysRevLett.101.026403} 
Phonon frequencies were calculated on either a 4x4x2 (nonmagnetic), 2x2x2 (striped \& SDW) or 
4x4x1 (striped) q-mesh and then interpolated along several symmetry directions. The resulting 
phonon frequencies and eigenvectors were used to calculate the dynamical structure factor along 
the selected scan directions. The dynamical structure factor, which is proportional to the x-ray
scattering intensity, is given in Eqn. \ref{sf_all}.\cite{lovesey1984,JSSJ.58.5.205,RPP.63.171} In these equations,
$W_{d}\left(\mathbf{Q}\right)$ is the Debye-Waller factor, $n_{j}\left(\mathbf{q}\right)$ is the Bose-Einstein distribution,
$f_{d}\left( \mathbf{Q}\right)$ is the x-ray form factor,  and $\mathbf{\sigma}_{\mathbf{d}}^{j}(\mathbf{q})$ is the eigenvector corresponding to the 
normalized motion of atom d in the $\mathrm{j^{th}}$ phonon branch. While the DFT calculation does not include temperature dependence, the Bose-Einstein distribution was set to 300K.

\begin{subequations}
\label{sf_all}
\begin{align}
S_{j}\left(\mathbf{q},\omega\right) & = \frac{\left|H_{\mathbf{q}}^{j}\left(\mathbf{Q}\right)\right|^{2}}{2\omega_{j}\left(\mathbf{q}\right)}\left(1+n_{j}\left(\mathbf{q}\right)\right)\delta\left\{ \omega-\omega_j\left(\mathbf{q}\right)\right\} \label{eq: structure factor 1}\\
H_{\mathbf{q}}^{j}\left(\mathbf{Q}\right) & = \sum_{d}\frac{f_{d}\left( \mathbf{Q} \right)}{\sqrt{M}_{d}}\exp\left(-W_{d}\left(\mathbf{Q}\right)+i\mathbf{Q}\cdot\mathbf{d}\right)\left\{ \mathbf{Q\cdot\mathbf{\sigma_{d}^{j}\left(\mathbf{q}\right)}}\right\} \label{eq: structure factor 2}\\
W_{d}\left(\mathbf{Q}\right) & = \frac{\hbar}{4M_{d}\Omega_{BZ}} \int_{\Omega_{BZ}} \sum_{j} \frac{| \mathbf{Q} \cdot \mathbf{\sigma}_{dj} |}{\omega_{j}} \left< 2 n_j \left( \mathbf{q} \right) + 1\right> \label{eq: DW factor}
\end{align}
\end{subequations}

Both the x-ray form factor and the Debye-Waller factor decrease intensity of the phonon excitations with increasing Q.
 The preferred approach for computer applications are numerical approximations to the x-ray form factor. The x-ray form factor has been parameterized by Waazmaier and Kirfel as the sum of five Gaussians plus a constant term.\cite{AC.A51.416} The Debye-Waller factor, calculated using Eqn. \ref{eq: DW factor}, can be thought of as the mean-squared value of the displacement each
atom dotted with $\mathbf{Q}$. The volume integral was calculated using 
the tetrahedron method \cite{CPC.157.17,PhysRevB.49.16223} on a 16x16x8 (nonmagnetic), 
16x16x4 (checkerboard) and 12x12x12 (SDW \& striped) Monkhorst-Pack q-point grid.\cite{PhysRevB.13.5188}
 To second order, the integral over the tetrahedron is simply 
the function evaluated at the center point multiplied by the volume. To avoid repeating the calculation 
for each value of Q, the nine potential components of the phonon eigenvector were stored and the dot 
product calculated later. 


The delta function in $\omega$ was convoluted with the elastic scattering width of Plexiglas 
 measured on analyzer 5. The pseudo-Voigt function fits well to the center of each peak, but small 
discrepancies exist in the tail. To minimize this effect a discrete linear convolution between 
the raw experimental data and simulated delta functions (single point on a grid) was performed 
numerically.

In addition to the energy resolution, the diameter of the analyzer leads to a finite resolution in \textbf{Q}. 
Slightly different positions on the analyzer can be described by a radial component, 
determined from the size of the analyzer (10 cm) and the distance from the sample to the analyzer (9 m),
and an angular component covering the entire circle. Values of \textbf{Q} accepted by the analyzer can be written as a function of these two variables. 
5000 samples from a pseudo-random number generator gave sufficient precision for convolution of constant-\textbf{Q} scans with the \textbf{Q}-space resolution. 
Due to the lower required precision 1000 samples were used for each contour plot. 

Figure \ref{line scan} shows a constant-\textbf{Q} energy scan at  $\left(0,0,8.3\right)$ and $\left(0,0,8.5\right)$ at room temperature.
Experimental data are given by the green points and pseudo-Voigt fits by the solid black line. The default values for $\eta$ and
$\Gamma$ only account for the energy resolution. At (0,0,8.3) the fit on the acoustic mode at 5 meV, however, is adjusted by including $\eta$ and $\Gamma$ as variables. The fitted values of $\eta$ and $\Gamma$ are $0.48\pm0.07$ and $3.96\pm0.09$, respectively. Optical modes are present at 22, 27, and 34 meV. At (0,0,8.5) the fit on the acoustic mode at 8 meV, however, was adjusted by setting $\eta=1.0$ (Lorentzian function) and including $\Gamma$ as a variable. The fitted 
value of $\Gamma$ is $3.67\pm0.12$ meV. A nearby optical mode is present at 11 meV, along with three other modes at 22, 26 and 34 meV, respectively.        

\begin{figure}[htp]
\includegraphics[]{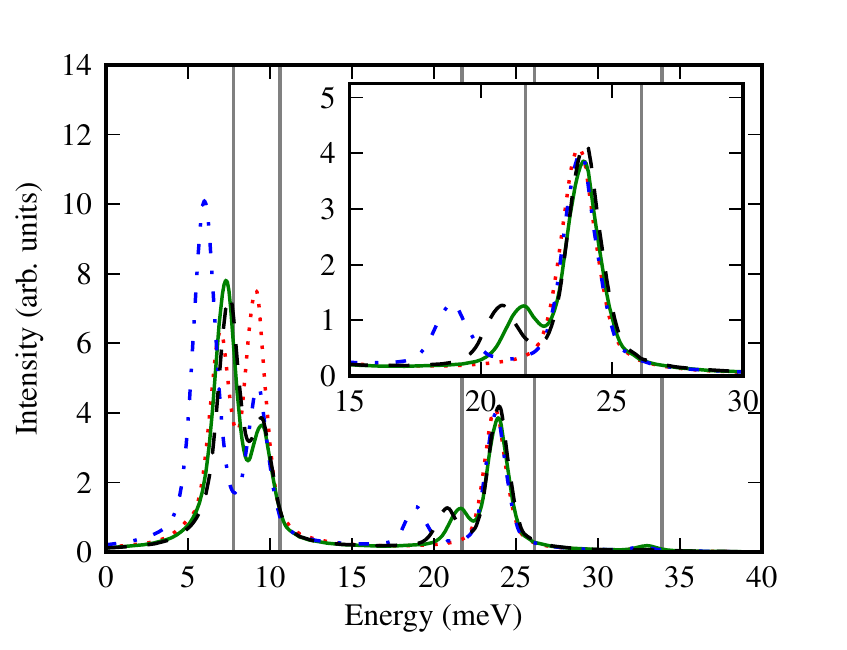}\tabularnewline
\caption{(color online) Dynamical structure factor calculation of constant-Q line scan at $Q=\left(0,0,8.5\right)$. The dotted red line corresponds to non-magnetic calculations of the 
dynamical structure factor.  The solid green line corresponds
to spin-polarized calculations imposing the SDW AFM ordering observed at lower temperatures.
 The black dashed line and blue dashed-dotted lines correspond to spin-polarized calculations 
with a striped (ferromagnetic along \textit{c}) and checkerboard ordering, respectively. The experimentally observed frequencies in Fig. \ref{line scan}b are shown with vertical grey lines.\label{line scan zoom}}
\end{figure}

\begin{figure}[htp]
\includegraphics{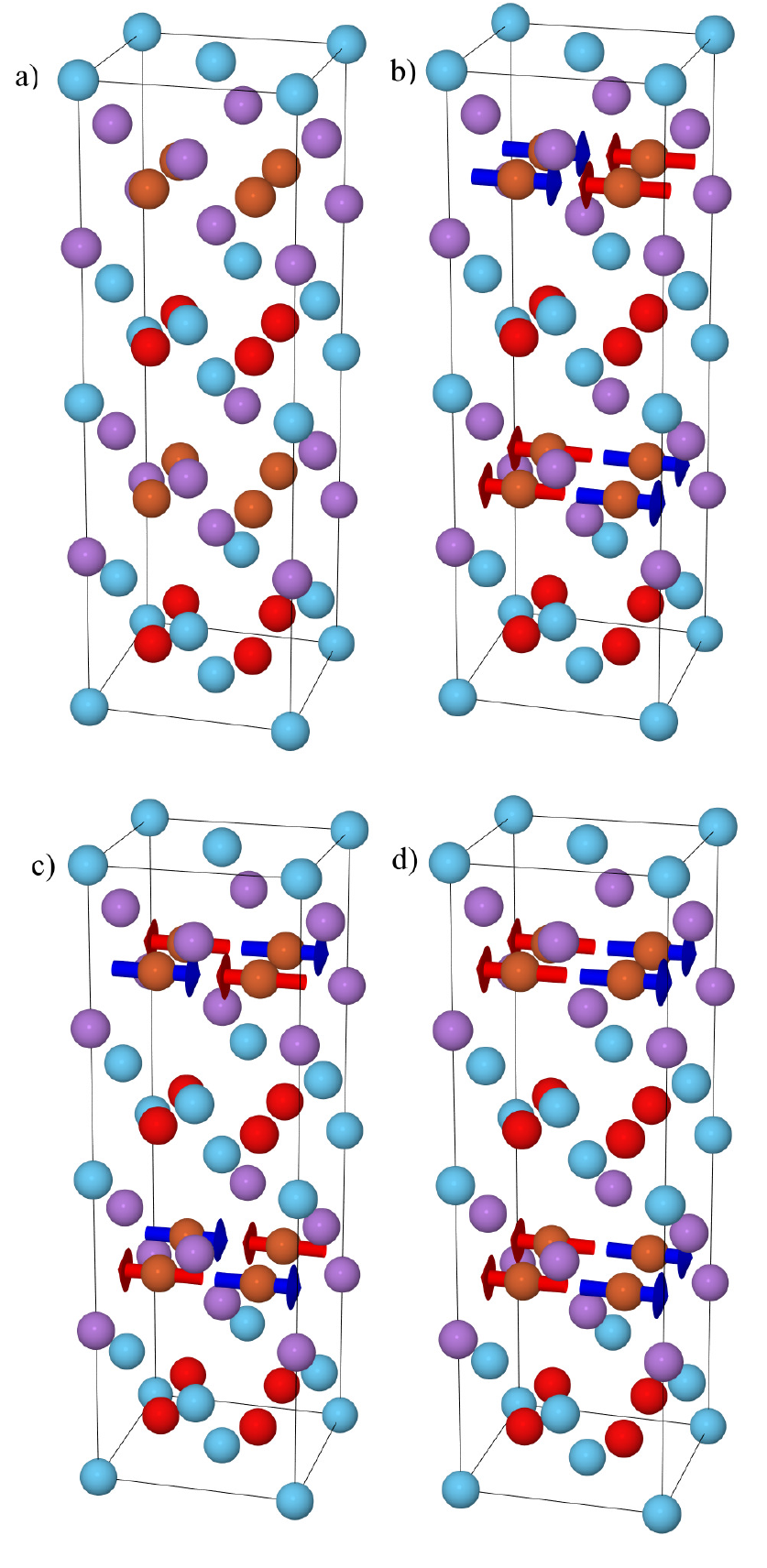}%
\caption{(color online) Different AFM order used in the calculations. La atoms are light blue, 
O atoms green, As atoms purple and Fe atoms brown. The red and blue arrows show up and 
down spin, respectively. a) expanded non-magnetic unit cell, b) experimentally observed SDW, c) 
Checkerboard ordering d) striped ordering aligned ferromagnetically along the \textit{c}-axis \label{lattice}}
\end{figure}

Fig. \ref{line scan zoom} shows several calculations of the dynamical structure factor at \textbf{Q}=$\left(0,0,8.5\right)$ which can be directly compared to Fig. \ref{line scan}b. The red dotted
line is a non-magnetic calculation. Frequencies for the acoustic and lowest optical 
modes are reasonable, but the calculated intensity of the optical mode is too high. Attempts to include the experimental 
uncertainty in \textbf{Q} could not reproduce the observed broadening of the acoustic mode, meaning it is not an artifact of Q-space resolution. The phonon excitation near 24 meV consists of two 
modes separated by 0.2 meV. At the zone boundary, lower and upper $A_{1g}$ modes consist of As and La motion, 
respectively, polarized along the \textit{c}-axis. At the zone center, these modes are mixed, each containing both La and As motion and the upper mode contributes 80\% of the structure factor. This result from the nonmagnetic calculation is 
inconsistent with the measurements, where these two modes are clearly split by 4 meV at (0,0,8.5). While the calculated 
frequencies agree with other published phonon dispersions,\cite{PhysRevLett.100.237003, PhysRevLett.101.026403} they are a few meV lower in energy than observed. Small changes in lattice parameters are not responsible, as an unphysical ~7\% reduction in the unit cell volume is required to stiffen this phonon mode in the nonmagnetic calculation to the observed value. While this discrepancy exposes limits on the accuracy of these DFT calculations, this should not detract from qualitative changes between calculations, such as the splitting of the $A_{1g}$ branches, that are also observed experimentally. At both values of Q, the 32 meV feature consists of both Fe and As motion, but the intensity is extremely weak.

In the spin-polarized calculation corresponding to the observed stripe AFM structure (Fig. \ref{lattice}b),  the effect of the magnetization on Fe is to strongly split these two branches at (0,0,8.5) with the ~21 meV excitation, 
containing As motion, lowering in energy by approximately 8.6\%. The ratio of intensities between the acoustic and nearby optical 
mode moves in the direction of, though slightly more than, what is observed experimentally. The 24 meV peak is primarily La motion. 
The intensity of the 32meV feature is 5.2 times stronger, in better agreement with experiment.

In order to better understand the importance of the specific magnetic order and size of the Fe moment on the lattice dynamics, two additional calculations
were performed in hypothetical magnetic structures. First is the ``checkerboard" magnetic structure, shown in Fig. \ref{lattice}c. It is a tetragonal space group, where Fe neighbors have opposite spins. This calculation converges to a solution 0.023 Ry higher in energy and 
energy and with an 18\% smaller magnetic moment per Fe atom.  The acoustic mode is slightly softer and has greater intensity. The ~21 meV excitation, 
containing As motion, is lower in energy by approximately 19.9\%. The intensity of the 32meV feature is 2.9 times stronger than in the nonmagnetic calculation.

Second is the CeFeAsO structure,\cite{NatMat.7.953} also referred to as ``striped,"  and shown in Fig. \ref{lattice}d. It is an orthorhombic space group with ferromagnetic coupling
of Fe moments along the \textit{c}-axis. The dynamical structure factor for this material is shown with black dashes in Fig. \ref{line scan zoom}. The frequency and intensity of the acoustic and
optical modes at 8 and 11 meV are nearly identical. Once again, the effect of the magnetization on Fe is to strongly split these branches, 
with the ~21 meV excitation lowering in energy by approximately 11.9\%. The intensity of the 32meV feature is half as strong as the nonmagnetic calculation.

Fig. \ref{sf 00L} compiles all of the experimental data and calculations of the different magnetic structures by showing several contour plots of the dynamical structure factor along $\left(0,0,L\right)$. Values of calculated intensities range from blue (low intensity) to red 
(high intensity), and have been multiplied by the energy to improve visibility of the optical modes. The white dots show the experimentally determined frequencies with the intensity shown by the size of the dot.  In each
case, calculations with a magnetic moment on the Fe show splitting between the two $A_{1g}$ modes near 24 meV. At the zone center, 
the upper mode softens by 3.7 (SDW) - 5.3 (checkerboard) \% and the lower mode softens by 9.1 (SDW) - 16.6 (checkerboard) \%. Calculated 
frequencies of these two modes are a few meV lower than observed. Comparing nonmagnetic and spin-polarized calculations, 
the frequency of the zone boundary the upper La-As mode is essentially unchanged ($<0.6\%$). The intensity of the lower mode is strongest near (0,0,9), and the intensity of 
the upper mode is strongest near (0,0,8). The frequency of the lower mode differs by 2 meV. The SDW calculation best matches the experimental frequency, but the checkerboard pattern best matches the observed splitting between these two modes. We note that the checkerboard ordering also introduces a pronounced softening of the longitudinal acoustic mode when compared to the non-magnetic calculation and other magnetically ordered structures. Finally, we point out that the intensity of the optical mode near 10 meV is highest in nonmagnetic calculation. Overall changes in the phonon frequencies and intensities indicate the complex and subtle effects that magnetic ordering has on the lattice dynamics.

Despite the changes introduced by magnetic order, all the spectra are qualitatively similar for different magnetic orders, and in better agreement with experiment compared with nonmagnetic calculations.  This might be understood to occur as a consequence of Fe moments still being present above $T_N$, though without long-range order.\cite{PRB.81.214407} Compared to nonmagnetic calculations, imposing an AFM ordering better describes phonons in LaFeAsO. Consequently, it is likely that the presence of Fe moments, ordered or not, affects the force constants. Considering only z-polarized phonon branches containing La and As motion significantly reduces the number of force constants that contribute. First, only the ``zz" term in the 3x3 force constant tensor can contribute, greatly simplifying comparisons between different magnetic unit cells. Fe and O are essentially stationary in the modes considered, meaning force constants between Fe-Fe,Fe-O and O-O atoms do not contribute. La-La and La-O force constants are essentially unchanged in each calculation, and the bond distance between La-Fe is large and the resulting force constant small. Therefore, we can limit ourselves to the ``zz" term for La-As, As-As, and Fe-As force constants. Of these, the Fe-As force constant is the largest by an order of magnitude. Even with these simplifications, there was no clear softening of any specific pair-wise force. This provides additional evidence for T. Yildirim's observation that changes in the phonon modes are due to a complicated renormalization rather than softening of a single pair-wise force. \cite{PhysicaC.469.425} In the non-magnetic calculation the on-site force constants, which are a sum of all pair-wise interactions, are (in eV/$\text{\AA}^2$) (11.2,11.2,8.9) for Fe and (10.5,10.5,9.6) for As. They show significant softening around 10-20\% with the introduction of magnetic order, in good agreement with T. Yildirium's work. Small differences on the order of 0.2 eV/$\text{\AA}^2$ or less in the on-site force constants are likely from slightly different lattice parameters chosen in our calculations.  

In summary, we have measured the phonon dispersion along (0,0,L) tetragonal phase of LaFeAsO at room temperature, well above the magnetic ordering temperature of 138K. Nonmagnetic calculations
fail to reproduce the observed splitting between two $A_{1g}$ phonon modes at 22 and 26 meV. Spin-polarized first-principles calculations imposing a number of hypothetical antiferromagnetic orders are qualitatively similar and in better agreement with the experimental results than non-spin-polarized calculations. The presence of Fe-spins are necessary to predict the observed spectrum above $T_N$, however the renormalization of the force constants is quite complex and cannot be reduced to a single pair-wise force constant.

\begin{acknowledgments}
Work at the Ames Laboratory was supported by the Department of 
Energy-Basic Energy Sciences under Contract 
No. DE-AC02-07CH11358.
Use of the Advanced Photon Source, an Office of Science User Facility 
operated for the U.S. Department of Energy (DOE) Office of Science by 
Argonne National Laboratory, was supported by the U.S. DOE under 
Contract No. DE-AC02-06CH11357. The construction of HERIX was partially supported by the NSF under Grant No. DMR-0115852.
\end{acknowledgments}

\bibliography{LaFeAsO_phonons}

\end{document}